\documentclass[reprint,prb,amsmath,amssymb,floatfix]{revtex4-1}

\usepackage{amsmath}
\usepackage{upgreek}
\usepackage{graphicx}
\usepackage{dcolumn}

\begin{document}

\title{Ultra-low threshold polariton condensation}

\author{Mark Steger}
\author{Brian Fluegel}
\author{Kirstin Alberi}
\author{Angelo Mascarenhas}

\affiliation{National Renewable Energy Laboratory, 15013 Denver West Parkway, Golden, CO 80401}

\author{David W. Snoke}

\affiliation{University of Pittsburgh, Department of Physics and Astronomy, 100 Allen Hall 3941 O'Hara St Pittsburgh PA 15260}

\author{Loren N. Pfeiffer}
\author{Ken West}

\affiliation{Department of Electrical Engineering, Princeton University, Princeton, New Jersey 08544, USA}

\date{\today}



\begin{abstract}
We demonstrate condensation of microcavity polaritons with a very sharp threshold occuring at two orders of magnitude lower pump intensity than previous demonstrations of condensation.  The long cavity-lifetime and trapping and pumping geometries are crucial to the realization of this low threshold.  Polariton condensation, or ``polariton lasing'' has long been proposed as a promising source of coherent light at lower threshold than traditional lasing, and these results suggest methods to bring this threshold even lower.
\end{abstract}



\maketitle



Early observations of microcavity polariton condensation\cite{Kasprzak2006,Balili2007} were novel tabletop demonstrations of quantum many body physics.  In contrast to the cold atom systems to date, polariton studies require modest optics and lasers and operate at liquid helium temperatures.  Being a superposition of a quantum well (QW) exciton and a cavity photon, microcavity exciton-polaritons inherit interactions and a light mass, meaning they can thermalize and still exhibit quantum phenomena at elevated temperatures.  The strong coupling of the exciton and photon can be achieved by placing QWs within a properly tuned optical cavity, much like the structure of the regularly-used VCSEL.  For a thorough review of microcavity polaritons see \cite{Deng2010}.

Indeed, the accessibility and robustness of the solid-state microcavity polariton has made it a promising candidate for technological applications such as optical switching\cite{Menon2010,Gao2012,Anton2012,Steger2012,Ballarini2013} and coherent light emission\cite{Deng2003,Nelsen2009,Schneider2013,Bhattacharya2013}.  However, the early reports of polariton Bose-Einstein condensation (BEC) were criticized because the short polariton lifetimes do not allow the gas to fully thermalize--these systems represent quasicondensates rather than thermodynamically equilibrated BEC.  Improvements in the mirrors integrated into these structures increased the cavity lifetimes from $\sim$ 1 ps to 15 ps\cite{Wertz2010}, and later to 135 ps \cite{Nelsen2013, Steger2013, Steger2015}.  Here we show that this improved lifetime and the proper selection of pumping and trapping geometry can dramatically reduce the critical pumping density to achieve condensation.

Polariton lasing occurs in the strong coupling regime.  At high carrier densities, the exciton-photon coupling saturates and the polariton states revert to exciton and photon states.  In this regime, the cavity enters traditional, rate-driven lasing rather than thermodynamic-driven condensation.  For a system with a lower condensation threshold, there may be a wider range of output powers below the transition to standard lasing.  In early condensation results\cite{Kasprzak2006, Balili2007}, the condensate exhibited initial spectral narrowing upon condensation, but rapidly broadened with increasing density as the system approached saturation.  With a lower threshold, our polariton laser exhibits a minimal broadening over orders of magnitude of density change.


To put our condensation threshold into context, we compare our current results to previous polariton condensation demonstrations in Table \ref{tab:review}.  We consider similar systems to our GaAs microcavity polariton with experiments conducted at moderate cryogenic temperatures (4-20 K) and excited nonresonantly with a laser tuned to create free carriers on the order of 100 meV above the lower polariton state.  Typically, these experiments use either a continuous wave (CW) or quasi-CW excitation, so we compare condensation thresholds in terms of the instantaneous pump power and intensity at the condensate.

\begin{table*}[htbp]
\small
\centering
\caption{\bf Literature polariton lasing thresholds}
\begin{tabular}{cccccccc}
\hline
Source 
& \begin{tabular}{c} Lifetime\\(ps) \end{tabular}
& \begin{tabular}{c} Duty\\cycle \end{tabular}
& \begin{tabular}{c}Instantaneous\\power (mW)\end{tabular}
& \begin{tabular}{c}Instantaneous\\intensity (W/cm$^2$)$^{(1)}$\end{tabular}
& Trap type 
& \begin{tabular}{c}Condensate \\size ($\upmu$m)\end{tabular}
& \begin{tabular}{c}Pump \\diameter ($\upmu$m)\end{tabular} \\
\hline
\cite{Kasprzak2006} & 1 & 1\% & 16$^{(2)}$ & 1670$^{(2)}$ & defect (random) & 16 & 35 \\
\cite{Balili2007} & 4 & 2.40\% & 33 & 6800 & stress (harmonic) & 5 & 25 \\
\cite{Wertz2010} & 30 & 100\%$^{(3)}$ & 5.0 & 160000 & lithographic (wire) & - & 2 \\
\cite{Nelsen2013} & 270 & 100\% & 5.0 & 9900 & untrapped$^{(4)}$ & - & 8 \\
\cite{Nelsen2013} & 270 & 100\% & 30 & 60000 & self-trapped (1-d) & 10 & 8 \\
\cite{Liu2015} & 270 & 1\% & 100 & 200000 & \begin{tabular}{c}stress and \\optical (ring)\end{tabular} & 60 & 8 \\
This work & 270 & 7\% & 0.57 & 13 & stress (harmonic) & 5 & 75 \\ 
\end{tabular}
  \label{tab:review}
\raggedright
\small
\noindent \\ $^{(1)}$ Intensities approximated assuming uniform pump density.\\
$^{(2)}$ Source is not clear if this reported power density is average or instantaneous.  Here we assume the power is instantaneous. \\
$^{(3)}$  Source does not specify chopping, so we assume it is CW.\\
$^{(4)}$  Condensate coherence is low in the untrapped phase.
\end{table*}

Several results have been left out from this assessment because insufficient details were included.  For example \cite{Azzini2011} used a pulsed laser but did not indicate the repetition rate, making it impossible to extract the peak powers.  Ferrier et. al. \cite{Ferrier2011} explored interactions in terms of power relative to the threshold, but did not indicate the actual powers used.


The high-quality, long-lifetime sample used in this work is virtually identical to that studied in \cite{Nelsen2013,Steger2013,Steger2015,Liu2015}.  The stress-based trapping geometry was of the same method as \cite{Balili2007}.  We thinned the substrate to $\sim$ 50 $\upmu$m, to achieve a sharp ($\sim$ 50 $\upmu$m diameter) and deep ($\sim$ 2 meV) stress trap.  Starting at a large photonic detuning (-20 meV), we applied strain, redshifting the exciton until the center of the trap was nominally at resonant detuning.  While a very tight pump focus can be used to create a local barrier\cite{Liu2015}, in this work we defocused the pump spot to be slightly larger than the stress trap (pump diameter $\sim$ 75 $\upmu$m).  The laser was injected at an angle of 20$^{\circ}$.

The sample was cooled in a coldfinger-type Montana Instruments cryostat held at nominally 6 K.  Optical pumping and thermal conduction through the stressor assembly likely mean that the lattice in the vicinity of the condensate is warmer than this.  To mitigate warming from the optical pumping, the pump laser was chopped at 2 kHz with a 7\% duty cycle using a mechanical chopper.

Another distinguishing aspect of this study was the use of a stabilized CW TiSaph laser (M$^2$ Solstis) which offers greater frequency and intensity stability of the pump.  Since microcavity polaritons have dielectric mirrors, optical pumping efficiency is highly spectrally dependent.  Instability of the pump laser therefore can lead to a fluctuating density, which can severely broaden the condensation phase threshold.

\begin{figure}[htbp]
\centering
\fbox{\includegraphics[width=0.8\linewidth]{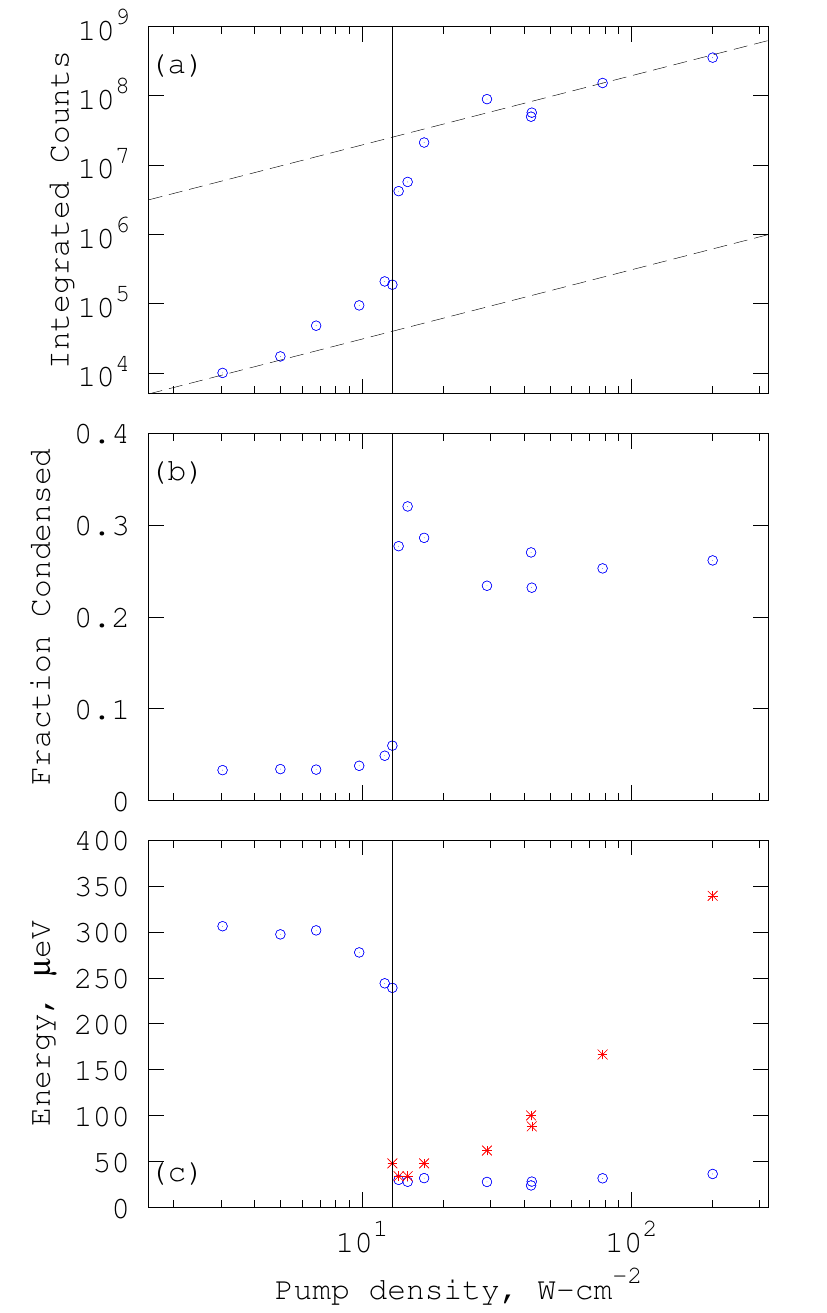}}
\caption{Pump-density dependence of the trapped polariton gas. Frame (a) shows the integrated counts of the trapped polaritons.  See the text for the method of integrating this density.  The dashed lines are guides to the eye for linear proportionality, and the dashed-dotted vertical line indicates the BEC phase transition density. The fraction of trapped polaritons in the most-occupied state is presented in frame (b).  Frame (c): blue circles: the spectral FWHM of the trapped polaritons (i.e. an indicator of their coherence) is shown to dramatically drop above threshold; red asterisks: the blueshift of the condensate relative to the low density ground state stays well below the 6 meV Rabi coupling strength, indicating that the system is still in strong coupling.}
\label{fig:analysis}
\end{figure}

After the optimal excitation wavelength (nominally 730 nm) was determined for the stressed sample position, we conducted a power series study, tracking the luminescence spectrum from the lower polaritons accumulated in the trap. Figure \ref{fig:analysis} shows our main results.

Figure \ref{fig:analysis} (a) shows the nonlinear jump in the total integrated luminescence of the trapped polaritons.  This sum of intensity counts was only conducted over the low-energy luminescence to exclude PL from outside the trap, which dominates the spectrum below the condensation density threshold.  The dashed lines are guides to the eye to highlight the linear relationships well below and well above the condensation threshold.  Frame (b) qualitatively indicates the fraction of particles in the most occupied state.  Below threshold, the most occupied state is not the ground state, due to bottleneck and thermalization effects.  Above threshold, this fraction underestimates the condensed fraction because it includes only the most intense spectral count rather than the sum over the coherent state.  However, together frames (a) and (b) clearly show that above threshold, polaritons are efficiently sucked into the condensate, with a relative reduction in the density of polaritons existing outside the trap or even in high energy states within the trap.

The blue circles in Fig.~\ref{fig:analysis} (c) show the dramatic spectral narrowing, measured by full width at half maximum (FWHM), of the polariton population within the trap.  In contrast to previous results (\cite{Kasprzak2006,Balili2007}), the linewidth of the condensed state stays pinned near our spectral resolution limit for more than an order of magnitude increase in pump power.  These earlier works showed an immediate increase in the spectral broadening of the condensate above threshold.

This can be clearly distinguished from traditional lasing because it occurs spectrally at the lower polariton energy, as shown by the red stars in Fig.~\ref{fig:analysis} (c).  If strong coupling were saturated, the lasing should occur at the cavity mode, which is 6 meV higher in energy than the lower polariton.  At most we observe $\sim$ 0.35 meV blueshift of the condensate energy over this extended pump power series study.

Following the work of Tassone and Yamamoto \cite{Tassone1999} we can estimate an upper limit of the density of excitons in the vicinity of the condensate.  The mean-field shift due to the interaction of excitons in QWs similar to ours is estimated to be $\sim 6$ $\upmu$eV/$\upmu$m$^2$\cite{Sun2015}.  Traditionally, the interaction strength of polaritons is assumed to scale according to their excitonic fraction, meaning that polaritons at resonance should interact with bare excitons with half this strength.  If we assume that the interactions with the bare excitons dominates the mean-field shift of the polariton states, we can express the spectral shift of this state to be $\Delta E = g n$ where $\Delta E$ is the blueshift of the polariton state, $g$ is the polariton-exciton interaction strength, and $n$ is the density of excitons.  At condensation onset we observe a 50 $\upmu$eV blueshift of the polariton ground state energy, which yields an estimate of 17 excitons/$\upmu$m$^2$.


The low threshold onset of polariton condensation can be qualitatively understood in terms of the longer polariton lifetime and the geometry of trapping and pumping.  Longer lifetimes means that the steady state will have commensurately higher populations for the same pumping powers.  Longer lifetimes also allow for better thermalization of the polariton gas.  However, as shown in \cite{Nelsen2013}, without trapping of the polaritons, they will travel ballistically away from the pump region, effectively reducing their lifetime and density.  Indeed, most untrapped polariton condensates exhibit reduced coherence \cite{Nelsen2013} or occupation of several states \cite{Wertz2010, Ferrier2011}.  While tightly focused pump spots can be effective for generating barriers on the fly \cite{Wertz2010, Ferrier2011, Liu2015, Sun2015, Sun2016}, it seems that the scattering into the low energy polariton states may be inhibited in the barrier region.  Here we have pumped a broad region with minimal renormalization and allowed the polaritons to cool and accumulate in the trap.

Whether lasing is achieved by optical or electrical pumping, the relevant parameter is the density of carriers at which coherence onsets.  Polariton lasing has been demonstrated to produce coherent light below the traditional lasing threshold, and here we have pushed the condensation onset to even lower absolute intensity by optimizing the polariton lifetime and trapping and pumping geometry.

The work performed at NREL is supported by the U.S. Department of Energy Office of Science, Basic Energy Sciences under DE-AC36-08GO28308; the work at Pitt was supported by the Army Research Office under Award No. W911NF-15-1-0466; and the sample growth at Princeton was funded by the Gordon and Betty Moore Foundation through the EPiQS initiative Grant GBMF4420, and by the National Science Foundation MRSEC Grant DMR-1420541.

We acknowledge D. Myers and J. Beaumariage for sample preparation and characterization.


\end{document}